# Spatially Nonuniform Oscillations in Ferrimagnets Based on an Atomistic Model

Xue Zhang[1†], Baofang Cai[2], Jie Ren[1], Zhengping Yuan[1], Zhengde Xu[1], Yumeng Yang[1,3], Gengchiau Liang[2], Zhifeng Zhu[1,3†]

[1]School of Information Science and Technology, ShanghaiTech University, Shanghai, China 201210

[2]Department of Electrical and Computer Engineering, National University of Singapore, Singapore 117576

[3]Shanghai Engineering Research Center of Energy Efficient and Custom AI IC, Shanghai, China 201210

**Abstract**

The ferrimagnets, such as $Gd_xFeCo_{(1-x)}$, can produce ultrafast magnetic switching and oscillation due to the strong exchange field. The two-sublattices macrospin model has been widely used to explain the experimental results. However, it fails in describing the spatial nonuniform magnetic dynamics which gives rises to many important phenomenons such as the domain walls and skyrmions. Here we develop the two-dimensional atomistic model and provide a torque analysis method to study the ferrimagnetic oscillation. Under the spin-transfer torque, the magnetization oscillates in the exchange mode or the flipped exchange mode. When the Gd composition is increased, the exchange mode firstly disappears, and then appears again as the magnetization compensation point is reached. We show that these results can only be explained by analyzing the spatial distribution of magnetization and effective fields. In particular, when the sample is small, a spatial nonuniform oscillation is also observed in the square film. Our work reveals the importance of spatial magnetic distributions in understanding the ferrimagnetic dynamics. The method developed in this paper provides an important tool to gain a deeper understanding of ferrimagnets and antiferromagnets. The observed ultrafast dynamics can also stimulate the development of THz oscillators.

**Introduction**

Terahertz (THz) frequency ranges from microwave to infrared [1], which has wide applications in the fields of biomedicine [2], materials science [3] and communication [4]. High frequencies can be produced by the current-induced oscillations in magnetic materials. In the most widely used ferromagnets (FMs), the frequency ranges from Megahertz (MHz) to Gigahertz (GHz) [5-7]. To generate and control higher frequency in the THz range, recent studies turn to the antiferromagnets (AFMs) [8-15], which consists of identical sublattices that are arranged antiparallelly through the strong exchange interaction. Theoretical studies have suggested that it is possible to control the AFM moments by the spin transfer torque (STT). The application of spin current on AFM leads to a THz precessing frequency. However, the material grain structure and the magnetoelastic effects make it more complicated to control the AFM moments [16, 17].

Similar to the AFM, the existence of strong exchange field in the ferrimagnet (FiM) allows it to generate high frequency in the THz range [18, 19]. However, the FiM is composed of different sublattices, which results in a symmetry breaking in the dynamic equation of the Neel vector. In addition, it exhibits finite magnetization, allowing the easy detection using the tunnel magnetoresistance effect (TMR). Furthermore, the ability to control the composition allows us to fabricate the FiM with different properties [20]. For example, the composition can be altered to reach the magnetization compensation ($x_{MC}$) or the angular momentum compensation ($x_{AMC}$)

[21, 22]. Previous studies have shown that the current induced magnetization oscillation in FiM can be classified as the FM mode with GHz frequency and exchange mode with THz oscillation [7, 19]. These theoretical studies describe the FiM using the two-sublattices macrospin model, where the magnetization dynamics is described by two coupled Landau-Lifshitz-Gilbert-Slonczewski (LLGS) equations [11, 19]. As a result, the two-sublattices macrospin model cannot capture the inhomogeneous magnetization dynamics such as the domain wall and the skyrmions, which can be significant as we have learned from the FM system [23]. The macrospin model has made a great contribution in describing the dynamics of FM. However, as a simplified model, the two-sublattices model lacks the spatial description of the FM system. Specifically, it is difficult to take into account the influence of neighboring atoms on the central atom. The same is true for FiM. Therefore, the spatial description in the two-dimensional atomistic model is particularly important for a more realistic description of the magnetization dynamics.

In this paper, we have developed a two-dimensional (2D) atomistic model to study the STT driven magnetization dynamics in the FiM, $(FeCo)_{1-x}Gd_x$, where $x$ denotes the Gd composition [24, 25]. We find that the direction of the charge current $J_c$ determines the chirality of magnetization oscillation. We propose a torque analysis method to understand this result. In addition, the variation of $x$ leads to different phase diagrams of magnetization oscillation. This can only be understood after taking the spatial nonuniform distribution of magnetic properties into consideration [26]. Furthermore, the size of system has a great influence on the stability of oscillation, which can be attributed to the nonuniform oscillation dynamics induced by the edge effect. These new results presented here reveal the necessity of studying the nonuniform magnetic properties in order to correctly understand the FiM dynamics.

**Methodology**

The 2D atomistic model is illustrated in Fig.1(a), where the Gd atoms are randomly distributed [27]. The FiM layer is then used as the free layer in the magnetic tunnel junction (MTJ) as shown in Fig. 1(b). The $J_c$ flows into the FiM layer and creates the STT acting on the magnetization. The magnetization dynamics in FiM is governed by the coupled LLGS equations [28],

$$\frac{\partial \mathbf{m}_i}{\partial t} = -\gamma_i \mathbf{m}_i \times \mathbf{H}_{\text{eff},i} + \alpha \mathbf{m}_i \times \frac{\partial \mathbf{m}_i}{\partial t} - \gamma_i B_{\text{D},i} \mathbf{m}_i \times (\mathbf{m}_i \times \mathbf{p}) \quad (1)$$

where $i$ denotes different sublattices. $\mathbf{p}$ is defined as the polarization of the pinned layer. The three terms on the right-hand side (RHS) represent the precession, the Gilbert damping and the damping-like STT, respectively. The effective field ($\mathbf{H}_{\text{eff}}$) consists of the exchange interaction and crystalline anisotropy. It is obtained from the Hamiltonian $\mathcal{H} = A\sum_i \mathbf{S}_i \cdot \mathbf{S}_{i+1} - K\sum_i (\mathbf{S}_i \cdot \hat{\mathbf{z}})^2$ with the exchange constant $A$ and the anisotropy constant $K$. As shown in Fig. 1(a), each atom is surrounded by four neighbors, resulting in three types of exchange interaction, i.e., $A_{\text{Gd-Gd}} = -1.26\times10^{-21}$ J, $A_{\text{FeCo-FeCo}} = -2.83\times10^{-21}$ J, $A_{\text{FeCo-Gd}} = 1.09\times10^{-21}$ J. The Hamiltonian expression of dipolar interaction is: $\mathcal{H}_{dipole} = -\frac{\mu_0}{4}\sum_{j\neq i}\frac{3(R_{ij}\cdot\mu_i)(R_{ij}\cdot\mu_j)}{R_{ij}^5} - \frac{\mu_i\cdot\mu_j}{R_{ij}^3}$, where $R_{ij}$ is the vector connecting spins $\mu_i$ and $\mu_j$. In the present sample with 100 atoms, the dipolar field that each atom receives from all other atoms is $10^3$ times smaller than the exchange field. Therefore, we ignore the dipolar interaction. $B_{\text{D},i} = \frac{\hbar}{2e}\frac{J_c\eta}{M_{s,i}t_{\text{FiM}}}$ represents the strength of STT, where $t_{\text{FiM}}$ is the thickness of the FiM layer, $\eta$ is the spin transfer efficiency, $M_s$ is the saturation magnetization. The magnetization dynamics study is performed by using a home-made code that numerically integrates the LLGS equations through the fourth–order Runge–Kutta methods (RKMs) [29]. The parameters are the same as that in [30], based on which we can determine $x_{\text{MC}} = 0.23$ and $x_{\text{AMC}} = 0.21$.

**Result and Discussion**

Fig. 1(c) shows the phase diagram of the current driven magnetization dynamics in the sample with $x = 0.1$. Under a negative $J_c$, the magnetization first switches (region 1), i.e., $\mathbf{m}_{\text{FeCo}}$ changes from $+\mathbf{z}$ to $-\mathbf{z}$ since $\mathbf{p}$ is opposite to the net magnetization. When $J_c$ is further increased, the magnetization of both atoms rotates in the counter-clockwise (CCW) direction at small angle (region 2), which is known as the exchange mode. For an even larger $J_c$, the effect of spin current overcomes the exchange interaction, resulting in the rotation of $\mathbf{m}_{\text{Gd}}$ in the sphere with $m_{z,\text{Gd}} < 0$ (cf. region 3), and we call it the flipped exchange mode. In this region, the atoms rotate in circles with different areas. However, since the atoms still experience strong exchange interaction, their oscillation frequencies are identical, which indicates that the magnetization in the larger circle has larger linear speed. Finally, when $J_c$ is further increased, both $\mathbf{m}_{\text{FeCo}}$ and $\mathbf{m}_{\text{Gd}}$ are aligned to the direction of $\mathbf{p}$, i.e., $-\mathbf{z}$ direction.

Similarly, when the positive $J_c$ is applied, both $\mathbf{m}_{\text{FeCo}}$ and $\mathbf{m}_{\text{Gd}}$ rotate in the clockwise (CW) direction that is opposite to the one under negative $J_c$. At larger positive $J_c$, the system enters the flipped exchange mode and finally both $\mathbf{m}_{\text{FeCo}}$ and $\mathbf{m}_{\text{Gd}}$ align along $\mathbf{p}$ in the $+\mathbf{z}$ direction. However, in the samples with a larger $x$, a different phase diagram is observed. As shown in Fig. 1(d) for the sample with $x = 0.15$, a negative $J_c$ first switches the magnetization, which is the same as the $x = 0.1$ sample. However, when $J_c$ is further increased, the system directly enters the flipped exchange mode. In this case, the exchange mode, where $\mathbf{m}_{FeCo}$ and $\mathbf{m}_{Gd}$ rotate in the opposite direction, does not exist anymore. The disappearance of exchange mode as a function of $x$ has not been reported before, and it cannot be explained using the two-sublattices macrospin model as discussed below.

Before studying the reason for the different phase diagrams as a function of $x$, we firstly provide a torque analysis method to understand the ferrimagnetic oscillation. The torques experienced by each atom under the current can be understood more clearly by converting the LLG equation (Eq. 1) into the Landau-Lifshitz (LL) form as

$$\frac{\partial \mathbf{m}_i}{\partial t}(1+\alpha^2) = -\gamma_i \mathbf{m}_i \times \mathbf{H}_{\text{eff},i} - \gamma_i \alpha \mathbf{m}_i \times (\mathbf{m}_i \times \mathbf{H}_{\text{eff},i}) + \alpha \gamma_i B_D \mathbf{m}_i \times \mathbf{p} - \gamma_i B_D \mathbf{m}_i \times (\mathbf{m}_i \times \mathbf{p})$$

(2)

from this equation, we can see that the stable oscillation can be initiated when the Gilbert damping (the second term on the RHS) is balanced by the damping-like STT (the last term). As shown in Fig. 2(a), when $J_c$ is applied in the +z direction, the Gilbert damping [−$\mathbf{m}\times(\mathbf{m}\times\mathbf{H}_{\text{eff}})$] and the damping-like STT [−$\mathbf{m}\times(\mathbf{m}\times\mathbf{p})$] acting on the Gd atom are pointing to the opposite directions. In contrast, these two torques on the FeCo atom are pointing to the same direction. Therefore, the magnetization oscillation in this case is initiated by the Gd atom, and then the FeCo atom is dragged into oscillation via the exchange interaction. Since the oscillation is initiated by the Gd atom, we can then determine the rotation direction by analyzing the precession torques experienced by Gd, i.e., the first and third terms on the RHS of Eq. (2). As shown in Fig. 2(a), both −$\mathbf{m}\times\mathbf{H}_{\text{eff}}$ and $\mathbf{m}\times\mathbf{p}$ are pointing to the same direction, resulting in the CW rotation when one looks from the top. This explains the magnetization oscillation and the rotation direction for the positive $J_c$ region in Fig. 1(c). Similarly, we can analyze the magnetization oscillation in the system where both $\mathbf{m}_{FeCo}$ and $\mathbf{J}_c$ are pointing to the $-\mathbf{z}$ direction. As shown in Fig. 2(b), the oscillation is still initiated by the Gd atom, on which the Gilbert damping and the damping-like STT are balanced. However, the atoms rotate in the CCW direction as a result of the precession torque. Therefore, the torque analysis method presented here agrees with the numerical phase diagram presented in Fig. 1(c), and we can conclude that in the sample with a fixed $x$, the magnetic oscillation (i.e., balance of torque) and rotation direction are determined by the same atom (Gd in this case) which is not related to the direction of $\mathbf{J}_c$ or the state of magnetization.

Based on the torque analysis method, we find that the steady oscillation only occurs when the Gilbert damping is balanced by the damping-like STT. Therefore, the oscillation mode is determined by the magnitude of $\mathbf{H}_{eff}$ and $\mathbf{J}_c$. For example, in the sample with a small $x = 0.1$, the Gilbert damping can be balanced by the damping-like STT at $J_c = +5\times10^{11}$ A/m$^2$, allowing the magnetic oscillation in the exchange mode [marked as the star in Fig. 1(c)]. In contrast, when $x$ is increased, the amount of Gd atoms is increased, resulting in more Gd-Gd interactions. Since $A_{Gd-Gd}$ is larger than $A_{FeCo-Gd}$, $\mathbf{H}_{eff,Gd}$ becomes larger. Therefore, when the current maintains at $J_c = +5\times10^{11}$ A/m$^2$, the Gd atom in the sample with increased $x$ can no longer maintain the torque balance required for the oscillation in the exchange mode [marked as the star in Fig. 1(d)]. However, at this point, the oscillation still occurs, but in the flipped exchange mode. Now we need to figure out why the torque balance can be achieved in this mode. Since the Gilbert damping is independent on the oscillation mode, it is the $\mathbf{H}_{eff,Gd}$ that has to be reduced. This can be realized in several ways. Firstly, some FeCo atoms around Gd can be flipped to reduce $\mathbf{H}_{eff,Gd}$. Assume $m_{z,Gd} < 0$ and the surrounding $m_{z,FeCo} > 0$, the corresponding

$H_{ex,Gd}$ points to −z direction, which combines with $H_{an,Gd}$ and the resulting $H_{eff,Gd}$ is too large to be balanced by the damping like STT. When some FeCo atoms are flipped to $m_{z,FeCo} < 0$, the exchange fields produced by these atoms change to +z direction. This reduces the $H_{ex,Gd}$ along the −z direction, so that $H_{eff,Gd}$ and the damping like STT can be balanced to initiate the oscillation. Although the oscillation condition can be satisfied under this picture, it cannot explain the oscillation in the flipped exchange mode at $J_c = +5\times10^{11}$ A/m$^2$, i.e., the average $m_{z,Gd}$ is larger than 0 in this mode. Therefore, in addition to the flipping of some FeCo atoms, we can further suspect that some Gd atoms are also switched from $m_{z,Gd} < 0$ hemisphere to $m_{z,Gd} > 0$ hemisphere to assist the oscillation. For example, when $m_{z,Gd} < 0$, both $H_{an,Gd}$ and $H_{ex,Gd}$ point in the −z direction, the damping-like STT provided by $J_c$ has to overcome both of them to initiate the oscillation. In contrast, if Gd atoms are flipped to $m_{z,Gd} > 0$, $H_{an,Gd}$ changes to +z direction, which assists the damping-like STT to balance with $H_{ex,Gd}$. Therefore, the results shown Fig. 1(d), i.e., the system oscillates in the flipped exchange mode instead of the exchange mode when x is increased to x = 0.15 at $J_c = +5\times10^{11}$ A/m$^2$, can be explained by combining several mechanisms. Furthermore, these explanations point out that it is necessary to take the complicated spatial magnetic information [31] into consideration, which cannot be captured by the macrospin model and one has to resort to the atomistic model.

To verify our explanations, we then look into the effect of spatial distribution on magnetization and effective fields. In Fig. 3(a), the Gd atoms are marked as the blue spheres in the sample with x = 0.15. The rest are the FeCo atoms. Under $J_c = +5\times10^{11}$ A/m$^2$, all the atoms oscillate and the average effect exhibits as the flipped exchange mode which corresponds to Fig. 1(d). $m_z$ of each atom is denoted in the color bar with red and blue represents +z and −z, respectively. It can be clearly seen that the magnetization of some atoms has been flipped to the opposite state, i.e., some FeCo and Gd atoms have been flipped to the $m_z < 0$ and $m_z > 0$ hemisphere, respectively. Furthermore, we plot the $H_{ex,z}$ experienced by each atom in Fig. 3(b). It can be seen that the exchange field near the Gd atom are generally small, which forms a boundary between the Gd and FeCo atoms. The apparent drop of the exchange field at the boundary separating Gd and FeCo atoms supports our explanations that $\mathbf{H}_{ex,Gd}$ is required to be reduced to maintain the oscillation, and this can be realized by flipping the magnetization of some Gd and FeCo atoms. In comparison, in the sample with $x = 0.1$ and $J_c = +5\times10^{11}$ A/m$^2$, the magnetization oscillates in the exchange mode as shown in Fig. 1(c).

In addition, some discontinuities appear at the boundary between the flipped exchange mode (region 3) and the region 4. At this boundary, we observed an unstable oscillation, which does not occur in the sample with $x$ larger than 0.15. This unstable oscillation is manifested as the back and forth fluctuation of the angle between $\mathbf{m}_{Gd}$ and the +z axis. We attribute these discontinuities to the unstable oscillation. At this boundary, since most Gd atoms are pointing to the hemisphere with $\mathbf{m}_{Gd} > 0$, the oscillation condition requires that $\mathbf{H}_{eff,Gd}$ should align to the –z direction to balance the torques. This can be achieved by either pulling $\mathbf{m}_{FeCo}$ to the +z axis or moving $\mathbf{m}_{Gd}$ away from the +z axis. In the sample with smaller $x$, $\mathbf{H}_{eff,FeCo}$ is larger, which makes all $\mathbf{m}_{FeCo}$ already aligned to the +z axis. Therefore, only the latter option is feasible. However, in this case, STT pulls $\mathbf{m}_{Gd}$ to the +z direction. Their competition leads to the back and forth movement of $\mathbf{m}_{Gd}$.

Fig. 3(c) shows the comparison of current range for the two oscillation modes as a function of $x$. The ratio in the y axis is calculated as the current range of the exchange mode over the entire oscillation range. When $x$ is small, both modes exist, and the current range of the exchange mode is around half as small as the flipped exchange mode. As $x$ increases, the ratio is gradually reduced. At $x = 0.15$, the exchange mode disappears and the ratio remains zero until $x = x_{MC}$. As explained in Fig. 3(a) and 3(b), this is attributed to the change of spatial magnetic properties when the amount of Gd is varied. Interestingly, as $x$ is further increased, the exchange mode appears again and the corresponding current range expands as a function of $x$. Noticing that this transition happens at $x = x_{MC}$, we then explain this result based on the change of the dominate magnetization when $x$ exceeds $x_{MC}$. When $x$ is smaller than $x_{MC}$, the dominate magnetization is $m_{FeCo}$, and the positive $J_c$ drives the magnetization into oscillation [see Fig. 1(c)]. Note that this is different from the scenario under negative $J_c$, where the magnetization switching happens first followed by the oscillation. However, when $x$ exceeds $x_{MC}$, the dominate magnetization changes from $m_{FeCo}$ to $m_{Gd}$. In this case, the characteristics of the positive and negative $J_c$ swaps, i.e., under the positive $J_c$, the magnetization is firstly switched and then oscillating, whereas it directly enters oscillation under the negative $J_c$. The phase diagram for $x > x_{MC}$ is schematically illustrated as the insert of Fig. 3(c). As a result, the corresponding effective fields of FeCo and Gd atoms are also changed. For example, under $J_c = +5 \times 10^{11}$ A/m$^2$, FeCo points to the −**z** direction whereas Gd points to the +**z** direction. Using the torque analysis method presented in Fig. 2, we can find that the oscillation is now initiated by the FeCo atom, which is different from the sample with $x < x_{MC}$. To initiate the oscillation, the system resorts to the balance between $\mathbf{H}_{eff,FeCo}$ and the damping like STT acting on FeCo. In addition, in the samples with $0.15 < x < x_{MC}$, we attribute the disappearance of exchange mode to the increase of $\mathbf{H}_{eff,Gd}$ as a function of $x$. However, since the oscillation is determined by FeCo in the samples with $x > x_{MC}$, and the exchange interaction between FeCo-FeCo is stronger than FeCo-Gd, $\mathbf{H}_{eff,FeCo}$ decreases when the $x$ is increased. The reduction of $\mathbf{H}_{eff,FeCo}$ leads to the balance between the damping like STT and $\mathbf{H}_{eff,FeCo}$. Therefore, the system can oscillate in the exchange mode, without entering the flipped exchange mode. This explains the reappearance of the exchange mode when $x$ exceeds $x_{MC}$. In addition, $\mathbf{H}_{eff,FeCo}$ is further reduced as when $x$ is increased, resulting in a larger current range for the oscillation in exchange mode [cf. Fig.3(c)].

In the previous section, we discussed ferrimagnetic oscillation with a fixed sample size. As we have seen the importance of the spatial distribution, we finally study the effect of sample size on the ferrimagnetic oscillation. In this section, $x$ is set to 0.5 to avoid "non-integer Gd atoms" in different samples. For example, if we want to study the magnetization dynamics in different samples with $x$ fixed at 0.2, the number of Gd atoms will be 3.2 and 12.8 for the samples of 16 and 64 atoms, respectively. However, we have to set them as integer numbers in the code, e.g., 3 and 13. This variation in the number of atom will lead to an unfair comparison for samples with different size. This can be avoided by setting $x$ to 0.5. As a result, the resulting phase diagram for the systems studied in Fig. 4 is the same as the sample with $x = 0.1$ which is illustrated in Fig. 1(c). The relationship between frequency and $J_c$ at different sizes which is shown in Fig. 4(a), for the sample with 16 or 36 atoms, $f$ shows a step when $J_c$ is larger than $1.3 \times 10^{12}$ A/m$^2$. However, for larger samples, $f$ is independent on the size [see Fig. 4(b)]. It is worth noting that the nature of discontinuity shown in Fig. 4(a) is different from that in Fig. 1(c) which has been pointed out in the previous section. We have attributed the discontinuity in Fig. 1(c) to the back and forth oscillation of $\mathbf{m}_{Gd,z}$. In contrast, as shown in Fig. 4(c) and 4(d), the stable oscillations are confined in the **x-y** plane with $m_z$ remains the same. In addition, the frequency step occurs in the region of the flipped exchange mode (region 3) rather than at the boundary of regions 3 and 4. To understand these results, we study the oscillation trajectories of the sample with 16 atoms at $J_c = 1 \times 10^{12}$ A/m², where a uniform oscillation is observed [see Fig. 4(c)]. In contrast, for a larger $J_c = 1.3 \times 10^{12}$ A/m², the oscillation becomes nonuniform as shown in Fig. 4(d). The oscillation modes of both Fig. 4(c) and 4(d) belong to the flipped exchange mode. This nonuniform oscillation can be understood as the edge effect. In the system studied here, each center atom interacts with four neighboring atoms, where the edge atoms are only affected by two or three nearby atoms. When the number of edge atoms is larger than the center atoms, the averaged oscillation trajectory becomes nonuniform, resulting in the frequency step. For the systems studied here, this condition is only satisfied in samples with 16 and 36 atoms, whereas the number of center atoms will be dominating in samples with more than 36 atoms. These results also reveal that it is important to use the model that can capture the spatial dynamics during the study of magnetization switching or oscillation in a large sized sample.

**Conclusion**

In conclusion, the spatial dependent ferrimagnetic oscillation is studied using the two-dimensional atomistic model. As the composition of Gd in the sample is increased, it is found that the exchange mode firstly disappears, and then reappears after the magnetization compensation point is reached. By studying the spatial distribution of the magnetization and exchange field, we conclude that the spatial nonuniform magnetic properties have to be taken into consideration to correctly understand the magnetic dynamics in ferrimagnets. Furthermore, the oscillation dynamics is strongly affected by the sample size, which again emphasize the importance of the spatial information, which can only be described by the atomistic model. We also proposed a torque analysis method to gain a better understanding on the ferrimagnetic oscillation. The methodologies and results presented in this paper can greatly stimulate the study of the ultrafast ferrimagnetic or antiferromagnetic dynamics.

Corresponding Authors: †zhangxue2@shanghaitech.edu.cn, †zhuzhf@shanghaitech.edu.cn

**Acknowledgements**: X.Z, J.R, Z.Y., Z.X. Y.Y and Z.Z. acknowledge the support from the National Key R&D Program of China (Grant No. 2022YFB4401700), Shanghai Sailing Program (Grant No. 20YF1430400) and National Natural Science Foundation of China (Grants No. 12104301 and No. 62074099). B.C. and G.L. would thank the support by MOE-2017-T2-2-114, MOE-2019-T2-2-215 and FRC-A-8000194-01-00.

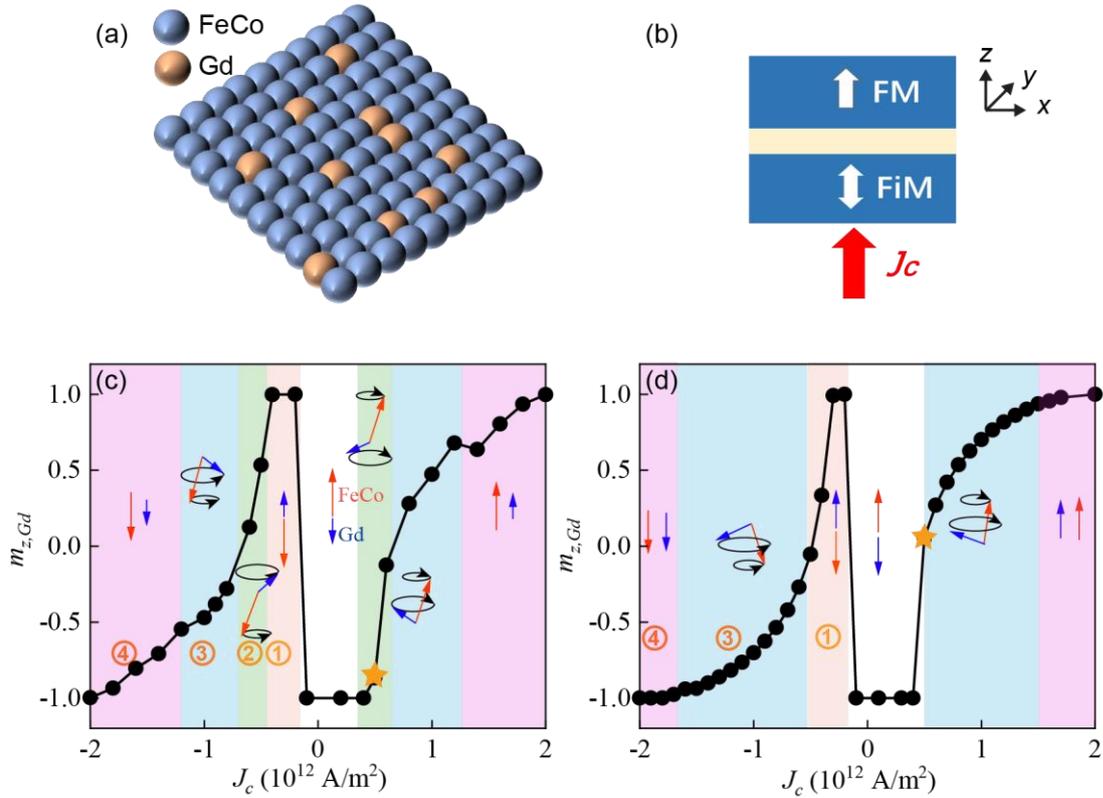

**Fig. 1.** Illustration of (a) the 2D atomistic model which consists of 100 atoms and (b) the device structure. Phase diagram of magnetization dynamics at (c) $x = 0.1$ and (d) $x = 0.15$. Red and blue arrows denote the magnetization direction of FeCo and Gd, respectively. $m$ is calculated by averaging the atoms of the same type. The point marked with the star represents $J_c = +5\times10^{11}$ A/m$^2$.

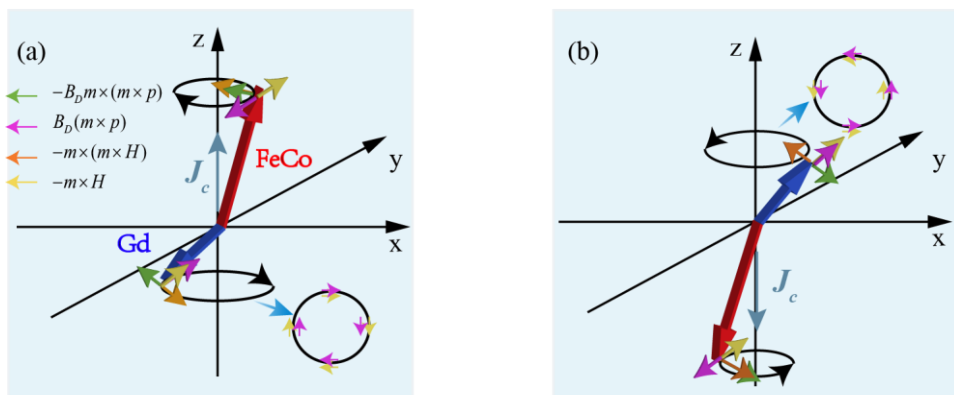

**Fig. 2.** Illustrations of the torques experienced by each atom when **m**$_{FeCo}$ and $J_c$ are pointing in the (a) +z and (b) −z directions.

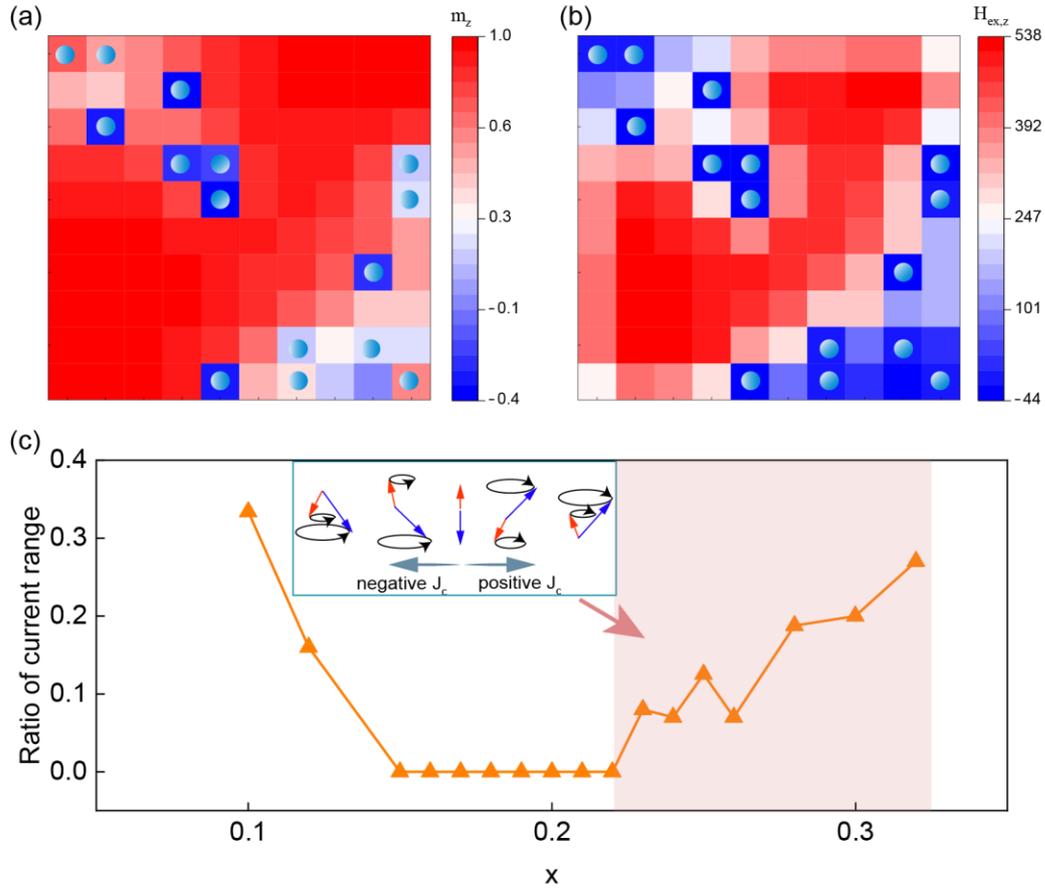

**Fig. 3.** Spatial distribution of (a) $m_z$ and (b) $H_{ex,z}$ in the sample with $x = 0.15$ and $J_c = +5 \times 10^{11}$ A/m². The squares with blue balls represent Gd atoms and others represent FeCo atoms. (c) Ratio of current range as a function of $x$. The insert illustrates the phase diagram for $x > x_{MC}$.

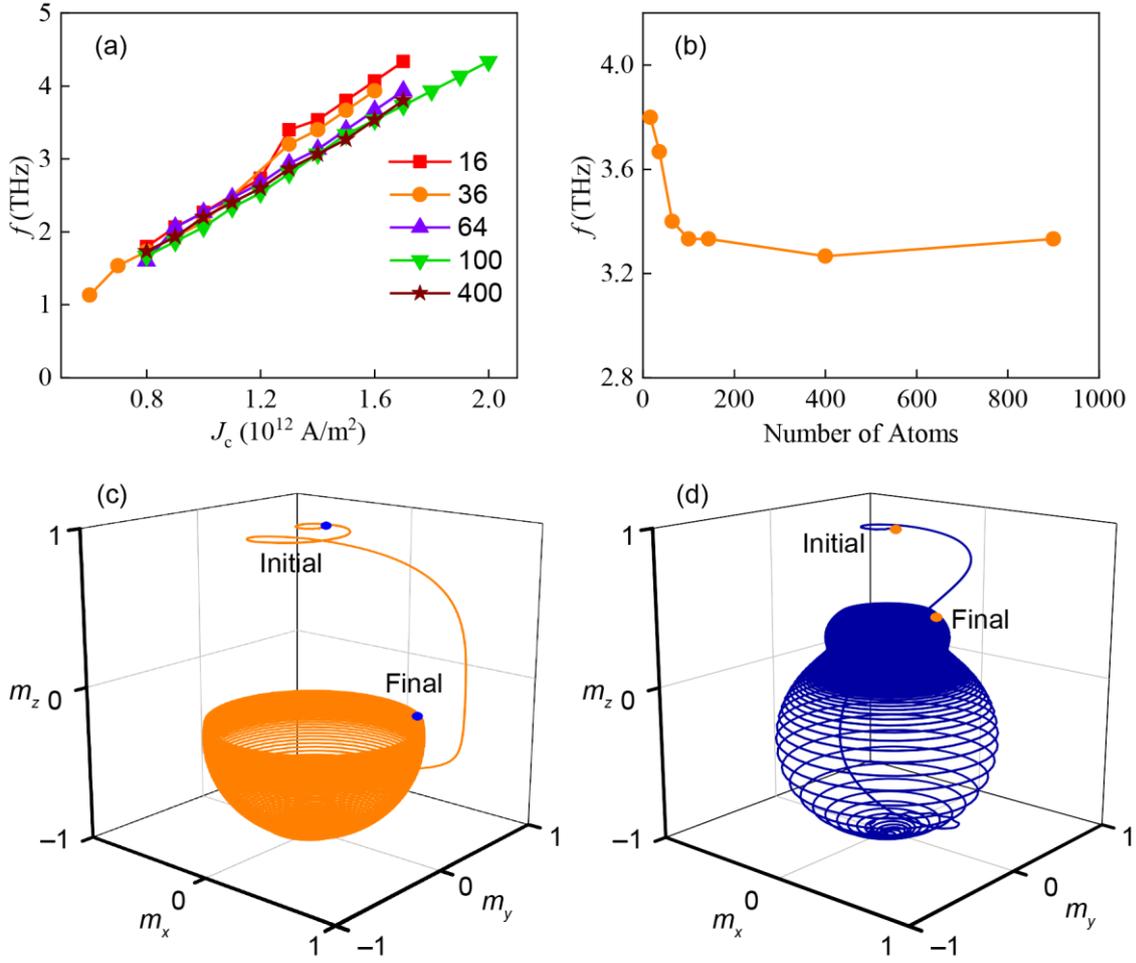

**Fig. 4.** (a) $f$ as a function of $J_c$ for samples with different sizes. The number of atoms is shown in the legend. (b) $f$ as a function of sample size with $J_c = 1.5\times10^{12}$ A/m². The average oscillation trajectory of the sample with 16 atoms at (c) $J_c = 1\times10^{12}$ A/m² and (d) $J_c = 1.3\times10^{12}$ A/m².